# Water-phospholipid interactions at the interface of lipid membranes: comparison of different force fields

Jianjun Jiang,[1,2] Weixin Li,[1,3] Liang Zhao,[1] Yuanyan Wu,[1] Peng Xiu,[4] Guoning Tang,[3] and Yusong Tu[1,a]

[1] College of Physical Science and Technology, Yangzhou University, Yangzhou 225009, China

[2] Department of Physics, Sanjiang College, Nanjing 210012, China

[3] College of Physics and Technology, Guangxi Normal University, Guilin 541004, China

[4] Department of Engineering Mechanics and Soft Matter Research Center, Zhejiang University, Hangzhou 310027, China

Water-phospholipid interactions at the lipid bilayer/water interfaces are of essential importance for the dynamics, stability and function of biological membrane, and are also strongly associated with numerous biological processes at the interfaces of lipid bilayers. Various force fields, such as the united-atom Berger force field, its two improved versions by Kukol and by Poger, and the all-atom Slipid force field developed recently, can be applied to simulating the structures of lipid bilayer, with their structural predictions in good agreement with experimental data. In this work, we show that despite the similarity in structural predictions of lipid bilayers, there are observable differences in formation of hydrogen bonds and the interaction energy profiles between water and phospholipid groups at the lipid bilayer/water interfaces, when four force fields for dipalmitoylphosphatidylcholine (DPPC) phospholipids are employed in molecular dynamics simulations. In particular, the Slipid force field yields more hydrogen bonds between water and phospholipids and more symmetrical interaction energy distributions for the two carboxylic groups on their respective acyl tails, compared to the Berger and its two improved force fields. These differences are mainly attributed to the different interfacial water distributions and ability to form hydrogen bonds between interfacial water and oxygen atoms of the DPPC lipids using different force fields. These results would be helpful in understanding the behaviors of water as well as its interaction with phospholipids at the lipid bilayer/water interfaces, and provide a guide for making the appropriate choice on the force field in simulations of lipid bilayers.

[a] Electronic address: ystu@yzu.edu.cn

# I. INTRODUCTION

Lipid membranes are the essential components of cells, and their structures and functions are closely related to the properties of water at the interface of biological membranes.[1,2] The membrane hydration has the significant influence on the phase transition temperature of membrane and lipid bilayer could become gel phase from fluid phase under the low hydration level of lipid bilayer.[3] The meticulous interaction between water molecules and membranes bring about the different structures of interfacial water from the bulk, and also bridges the interaction between lipid membranes.[4] Experiments indicate that there are different species of water molecules at the lipid bilayer/water interfaces, such as bound water, buried water and free water.[5] The results from neutron scattering on the interactions between hydration water and biological membrane show that the self-diffusion coefficient for the first layer of water at the membrane interface is five times smaller than that of bulk water.[6] Experimental results also show that there are about five tightly-bound water molecules per phospholipid molecule in phosphatidylcholine (PC),[7] and these water molecules are tightly hydrogen bonded (H-bonded) with the oxygen atoms of phospholipids,[8] and even with apolar methyl and methylene groups of phospholipids.[9]

Water behaviors at the lipid bilayer/water interfaces have been explored by employing the Molecular Dynamics (MD) simulations, which is a powerful tool for the understanding of many biological phenomena at the atomic or near-atomic level.[10] There have reported an increasing number of investigations on the water behaviors using the models of lipid bilayer/water interfaces.[11-15] A water molecule can simultaneously form H-bond with different oxygen atoms of single phospholipid or different phospholipids molecules, called as "water bridge", and thus the huge H-bond network can be formed at the membrane interface.[12] Hydrogen bonding structures and dynamics of the water molecules at the bilayers surface varies with the location of water molecules near the headgroups of phospholipids,[13] e.g., the deeply-buried water molecules usually have a longer H-bond lifetime.[14-16]

Many efforts have been made to develop more accurate force fields (FFs) for the simulations of lipid bilayers, in order to reproduce the properties of lipid bilayers and obtain better results with experimental measurements. So far, many types of FFs, such as all-atom, united-atom and coarse-grained FFs, have been proposed. For example, as a united-atom force field (FF), the Berger FF has been widely used,[17] also with its modification by Kukol[18] (Kukol FF) and the modification derived from GROMOS53A6 FF and Berger parameters by Poger[19] (Poger FF). As the all-atom FFs, CHARMM36 FF [20] and Slipid FF,[21] as well as Lipid14 FF,[22] have also been developed and updated. Most of these FFs can reproduce the structural properties of lipid bilayers such as bilayers thicknesses, the area per lipid, deuterium order parameters for the hydrophobic acyl chains of lipids and average electron density profiles of lipid bilayers, with a comparable accuracy to the available experimental data.[23, 24]

In this paper, we present a detailed comparison of water-phospholipid interactions at a fully hydrated bilayers interface based on the DPPC model using three united-atom FFs, namely Berger FF, Kukol FF, Poger FF, and one all-atom FF, namely Slipid FF. We find significant differences of the H-bonding network at the membrane interface, i.e., the energy distributions of H-bonds between interfacial water molecules and lipid molecules, although many structural properties of the lipid bilayers from these four FFs are in good agreement with experimental results. We hope our results can contribute to a better understanding of water-phospholipid interactions at the membrane interfaces and provide insights for choosing a more appropriate FF in biomembrane simulations.

## II. SYSTEMS AND METHODS

### A. Systems

The membrane systems are composed of 128 DPPC and 3810 water molecules (Fig. 1). The DPPC molecule formed by the glycerol backbone linked to the phosphatidylcholine moiety and attached to two acyl tails, including eight types of oxygen atoms (carboxylic oxygen, O33, O34, O22, O12, and ether oxygen, O31, O32, O21, O11) are highlighted in Fig. 1. Here we named O22 and O21 along with their attached carbon atoms as C=O2 group, O11 and O12 along with their attached carbon atoms as C=O1 group. The eight types of oxygen atoms belong to three groups, Phosphate, C=O1 and C=O2 groups, and therein, the carboxylic oxygen, O33, O34, O22 and O12, is double-bonded, and either oxygen, O31, O32, O21 and O11, is single-bonded.

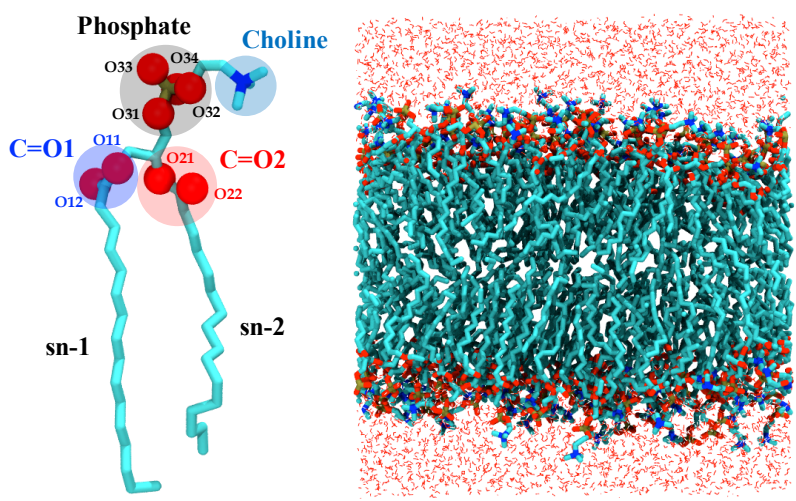

**Fig. 1.** Illustration of a DPPC lipid molecule and side view of the membrane system. Eight types of oxygen atoms: carboxylic oxygen, O33, O34, O22, O12, and ether oxygen, O31, O32, O21, O11 are shown in VDW balls. Four groups: Choline, Phosphate, C=O1 and C=O2 are in shaded areas.

As mentioned above, we performed the simulations using four lipid FFs, i.e., Kukol FF, Poger FF, Berger FF, and Slipid FF. The popular Berger FF [17] consists of a combination of parameters from GROMOS87 [25] and OPLS FF [26] with the modifications to the acyl chain by Berger and co-workers [17] and atomic partial charges from the calculations Chiu et al. [27] The Kukol FF is the improved versions developed from Berger and GROMOS53A6 FF with the modifications to carbonyl carbons which increase accuracy of membrane protein simulations. [28] The Poger FF is also developed from Berger and GROMOS53A6 FF with new atom types for the choline methyl and ester phosphate oxygen groups in phospholipids. [19] The all-atom Slipid FF derived from CHARMM36 FF, recalculates the parameters of lipid tails by using an even more precise initio method in an consistent manner with Amber FF. The simple point charge (SPC) water model [29] was employed in systems with Kukol, Poger, Berger, and the Slipid FFs.

### B. Molecular dynamics simulation and data analysis

All systems were performed with the Gromacs 4.5 software [31] under the isothermal-isobaric ensemble. The leap frog algorithm [32] was applied for the integration, and LINCS algorithm [33] was used for constraining bonds. Periodic boundary conditions were applied in all three dimensions. Four systems were maintained at the temperature of 323 K using the Berendsen algorithm with a coupling time of 0.1 ps. [34] The pressures in the lateral and normal direction were kept at 1 atm separately with the system coupled to a Berendsen barostat [34] with a coupling time of 2.0 ps. The electrostatic interactions were calculated with the particle mesh Ewald method, [35] with a real space cutoff of 1.2 nm. The Lennard-Jones interactions were truncated at 1.2 nm. Each system was run for 150 ns and the coordinates were saved every 1 ps. The trajectories during the last 20 ns were collected for analysis.

The hydrogen bond (H-bond) formed between water and DPPC molecules is defined according to the geometric criterion: the distance between two oxygen atoms is less than 0.35 nm and the O-H···O angle is less than $30°$. [36] The H-bond energy $E$ is defined as the sum of electrostatic energy and the Van der Waals potential energy, between water and corresponding phospholipid groups. [37-41]

### III. RESULT AND DISCUSSION
### A. Structural properties of the lipid bilayers under various force fields

We analyzed several important structural properties of lipid bilayers under different FFs, such as the average density profiles for various components and the deuterium order parameter of DPPC. As shown in Fig. 2, we present the average density profiles for water and different groups of the DPPC from the bilayer center, and the deuterium order parameter of hydrocarbon chains [19] of DPPC under one typical united-atom FF (Berger FF) and one all-atom FF (Slipid FF). The MD simulation results show that the results of the above two physical observables employing the Kukol and Poger FFs are closed to those under Berger FF, thus they are not shown in Fig. 2 for simplicity. Figures 2B and 2D show the results of the deuterium order parameter obtained from MD together with the experimental profiles. Note that the carbons are numbered consecutively starting with the carbonyl group of the acyl chain. One can find that the degree of order displays a trend of decrease along the chain toward the core of the bilayer. It can also be found that the structural properties of lipid bilayers gained from MD simulations with both the united-atom and all-atom FFs are within the experimental range and can well describe basic structural properties of lipid bilayers.[42]

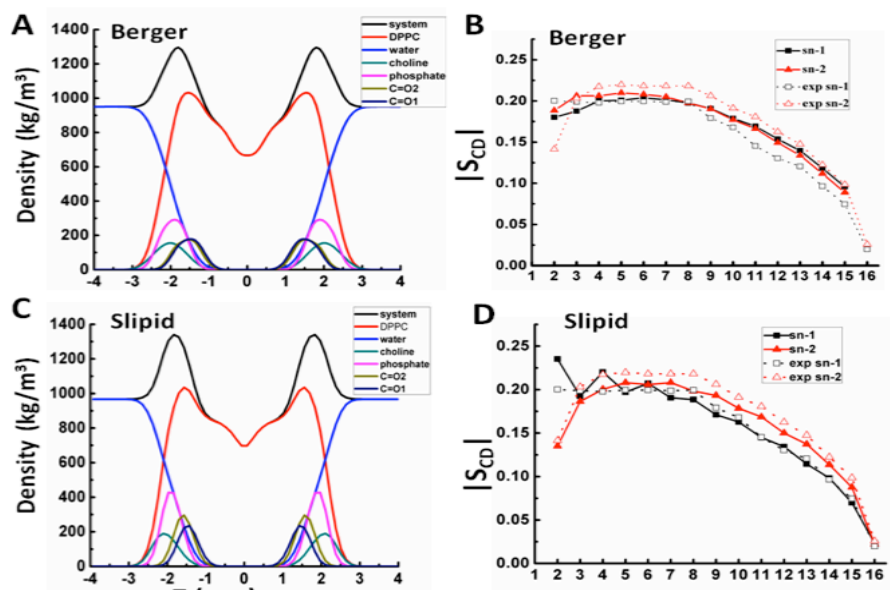

**Fig. 2.** The average density profiles for water and different groups of the DPPC as a function of distance from the bilayers center (z) and the deuterium order parameter of hydrocarbon chains of DPPC. (A) Density profiles using Berger FF and (C) Slipid FF. (B) The deuterium order parameter of DPPC using Berger FF and (D) Slipid FF. The experiment results [23, 24] are also shown for the easy comparison.

## B. The distributions of H-bond energy between water and phospholipids under different force fields

Water can be embedded into the hydrophilic region of lipid bilayer, and form the integrated H-bond network along with hydrophilic head matrix at the lipid bilayer/water interface. As shown in Fig. 3, we first take the Berger FF as an example to analyze these H-bond interactions between water molecules and head groups of phospholipids. Most of the H-bonds are formed between water molecules and those eight types of oxygen atoms of phospholipid head groups, mainly because of the large partial charges on the oxygen atoms and their exposure to water. The choline group is hydrophobic and difficult to form H-bonds with water.

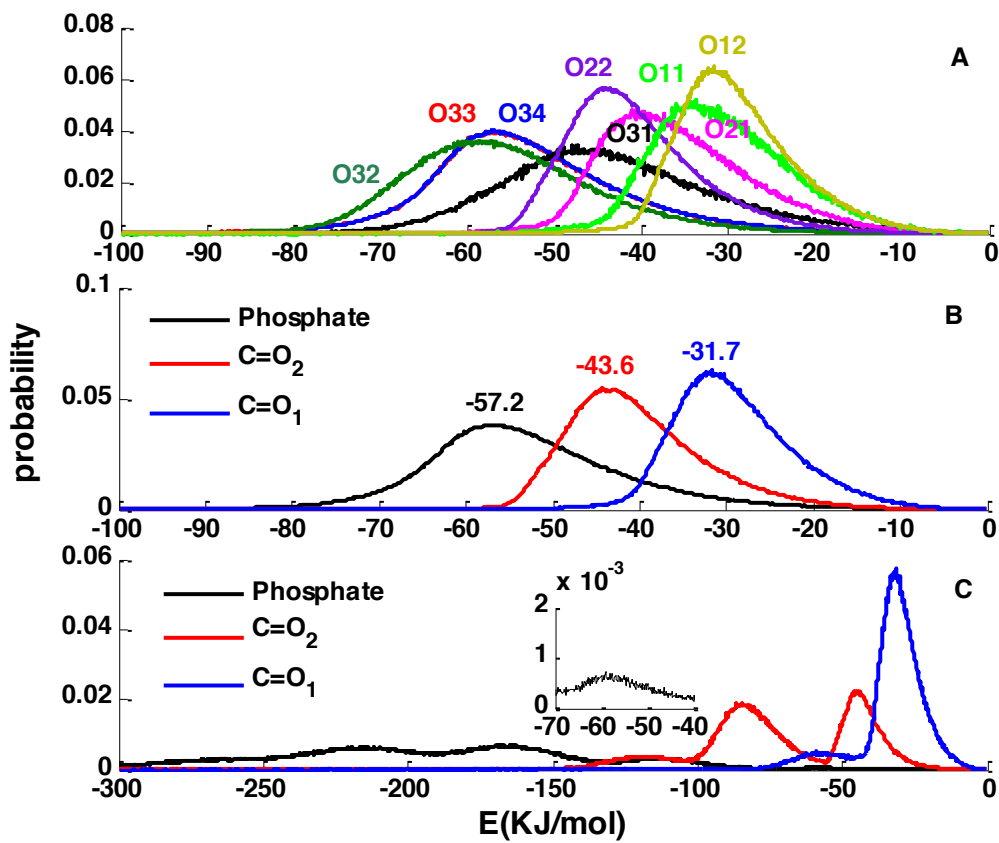

**Fig. 3.** (A) The H-bond energy distribution of each oxygen atom in DPPC. (B) The H-bond energy distributions between a water molecule and one group in DPPC. (C) The distribution of total H-bond energy for one group in DPPC. The inset in Fig. C shows the distribution of total H-bonding energy for Phosphate group in the vicinity of the first peak.

Fig. 3A presents the results of H-bond energy distributions of water molecules with specific oxygen atoms (see Fig. 1). We also classify the H-bond energy according to their corresponding groups, rather than the oxygen types. As shown in Fig. 3B, the three distributions of H-bond energy with the averaged value for Phosphate, C=O1 and C=O2 groups are obtained. It is noticeable that the average energies of H-bonds with the phosphate group are the largest value among all of three groups, indicating that H-bonds interaction between water and the phosphate group is stronger than two others. Finally, multiple water molecules can be simultaneously bonded to the same group, and thus the interaction between each water molecule and the group contributes to the total H-bonding energy. Fig. 3C presents the distributions of the total H-bonding energy of the lipid groups. For the phosphate group, there are four peaks corresponding to energies of -59, -115, -164, -217 kJ·mol$^{-1}$, approximately the integer multiple of the average energy (-57.2 kJ mol$^{-1}$) shown in Fig. 3B. Two higher peaks at the energy -164 and -217 kJ·mol$^{-1}$ indicate the higher probability of the simultaneous formation of three and four H-bonds with the phosphate group. Similarly, there is higher probability for two H-bonds with the C=O2 group and a single H-bonds with the C=O1 group.

Furthermore, we will compare the H-bond behaviors at the lipid bilayer/water interfaces under four FFs. In Fig. 4 we present the H-bonding energy distributions of water molecules with specific oxygen atoms under four FFs. Both united-atom FFs, Kukol FF and Poger FF are derived from Berger FF, and have the similar distributions and peak energies of the H-bonding energy of eight oxygen atoms. However, the H-bonding energy distributions with O11 under Kukol FF are slightly higher than that of O11 under Poger FF, caused by the modification of the carbon atoms type of DPPC in Kukol FF. Compared to Berger FF, the peak energy of H-bonding energy distributions with O32 slightly deviate from Kukol FF and Poger FF. Despite these small differences, it is noteworthy that the H-bonding energy distributions under the three united-atom FFs are indeed very similar to one another and don't display significant differences. However, significant differences can be found between the united-atom FFs and the all-atom FF (Slipid FF). As we can see from Fig. 4, although the H-bonding energy distributions of O33 and O34 from four FFs are completely overlapped, which reflects the symmetry of phosphate group, their peak energies from different FFs are different. The peak energies from Slipid FF are higher than those from the united-atom FF. Moreover, the difference between the H-bonding energy distributions of the double-bonded oxygen atom and that of single-bonded oxygen atom in the same group from Slipid FF is greater than that from the three united-atom FFs. This is mainly because that, as compared to Slipid FF, the difference of the partial charge between the double-bonded oxygen atom and the single-bonded oxygen atom in one group under united-atom FFs is smaller, as shown in Table 1.

Table 1. The partial charge of the eight kind of oxygen atoms for the four investigated systems.

| Force Field | O31 | O32 | O33 | O34 | O21 | O22 | O11 | O12 |
|---|---|---|---|---|---|---|---|---|
| Kukol | -0.70 | -0.80 | -0.80 | -0.80 | -0.70 | -0.70 | -0.70 | -0.60 |
| Poger | -0.70 | -0.80 | -0.80 | -0.80 | -0.70 | -0.70 | -0.70 | -0.60 |
| Berger | -0.70 | -0.80 | -0.80 | -0.80 | -0.70 | -0.70 | -0.70 | -0.60 |
| Slipid | -0.49 | -0.49 | -0.86 | -0.86 | -0.47 | -0.65 | -0.47 | -0.65 |

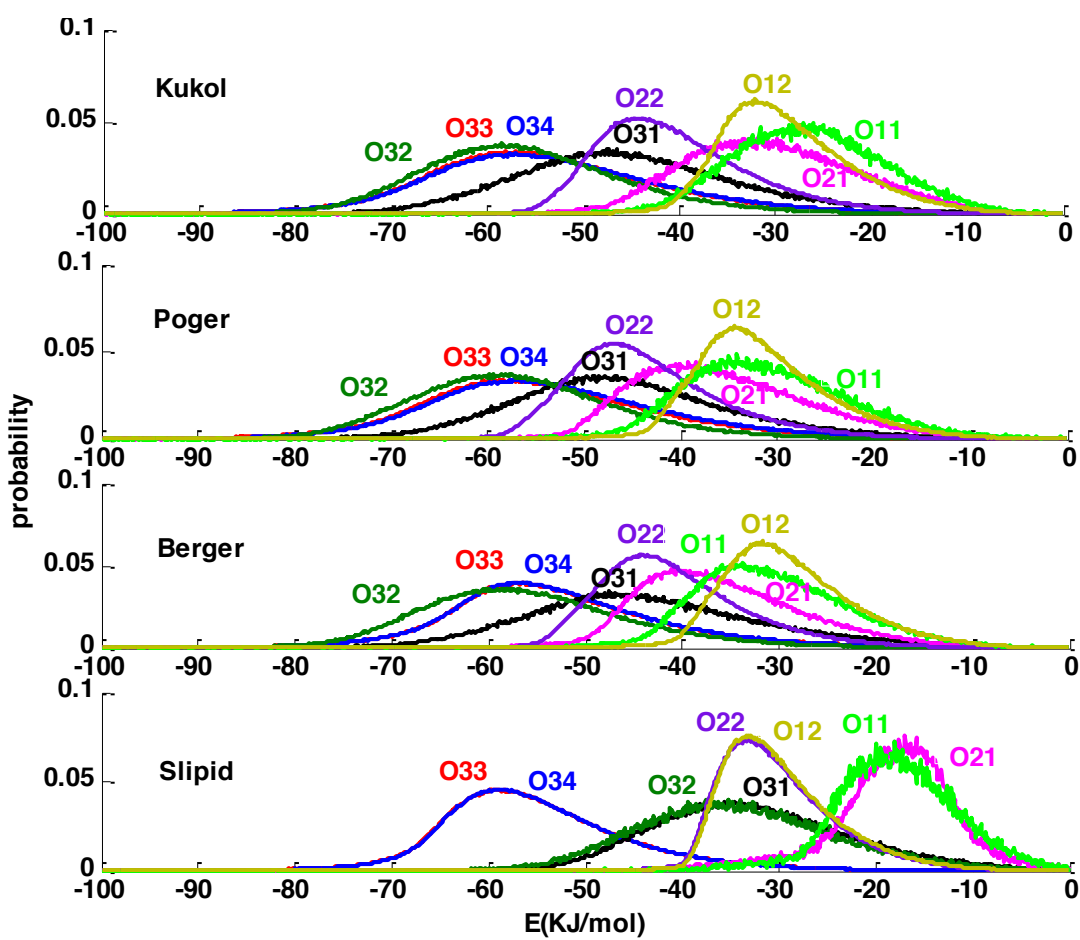

**Fig. 4.** The probability distributions of H-bond energy between waters and eight types of oxygen atoms under different FFs.

As shown in Fig. 5, we analyzed the total H-bond energy distributions between water and different lipid groups under four FFs. For the phosphate group, several peaks can be found in the total H-bonding energy distributions under four FFs. Kukol FF and Poger FF present three peaks, while Berger FF shows four peaks with a wider distribution. As an all-atom FF, the distribution of the total H-bond energy of phosphate group from Slipid FF also have four peaks, however, the third and fourth peaks have a higher probability and the distribution extends to -350 kJ mol$^{-1}$. For C=O2 group, three united-atom FFs yield two peaks with higher probability. However, only one peak with a sharp distribution appears under the all-atom Slipid FF. Obviously the hydrogen bonds of the glycerol groups under Slipid FF are weakened, as shown in Table 2. For C=O1 group, all the four FFs have two peaks. The height of the second peak is very low compared to that of the first one, indicating that there is a higher probability for the formation of H-bond with the C=O1 group.

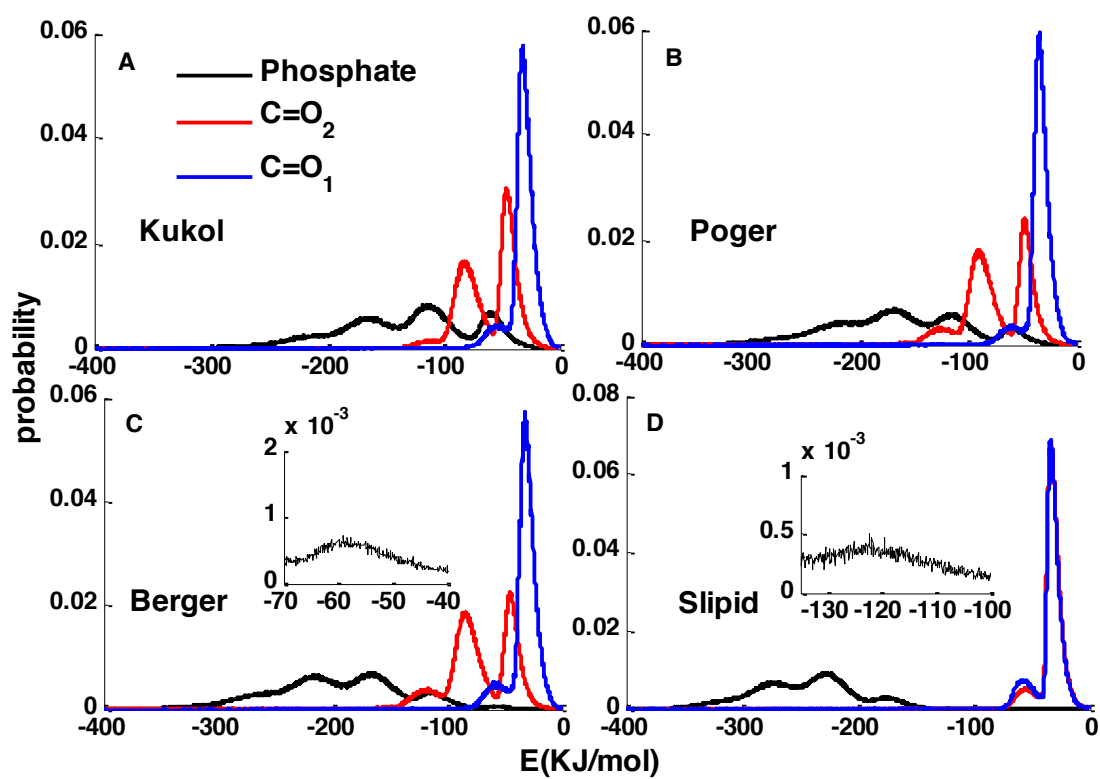

**Fig. 5.** The distributions of the total H-bond energy between waters and different lipid groups of DPPC under different FFs. The inset in Fig. C or D shows the distribution of total H-bond energy for Phosphate group in the vicinity of the first peak.

Table 2. The average numbers of H-bonds for different oxygens under different FFs.

| Force Field | O31 | O32 | O33 | O34 | O21 | O22 | O11 | O12 | Sum |
|---|---|---|---|---|---|---|---|---|---|
| **Kukol** | 0.1543 | 0.4163 | 0.9403 | 0.9570 | 0.0632 | 1.4006 | 0.0350 | 0.58274 | 4.4594 |
| **Poger** | 0.1699 | 0.4463 | 1.3151 | 1.1494 | 0.1231 | 1.5534 | 0.0362 | 0.4253 | 5.2187 |
| **Berger** | 0.1379 | 0.4590 | 1.5798 | 1.5690 | 0.2046 | 1.5456 | 0.0714 | 0.5195 | 6.0868 |
| **Slipid** | 0.1467 | 0.0446 | 2.1839 | 2.1846 | 0.0306 | 0.8471 | 0.0125 | 0.9151 | 6.3651 |

## C. The average numbers of H-bonds for different oxygen molecules under different force fields

Finally, we calculated the average numbers of H-bonds for different oxygen molecules at the lipid bilayer/water interfaces under the four FFs. As shown in Table 2, the H-bonds between water and DPPC are mostly formed around double bonded oxygens, namely O33, O34, O22, O12. The proportion of H-bonds in the four oxygens from Kukol FF is 87%, from Poger FF is 85%, from Berger FF is 86%, and from Slipid FF is 96%. It is interesting to find that although the partial charge for the eight kind of oxygen atoms from the above three United-Force fields are the same as shown in Table 1, there are obvious difference among the number of H-bonds under Berger, Kukol and Poger FF. To disclose the reason for this phenomenon, we calculated the average radial number density (ARND) of water around the eight kind of oxygen atoms from MD trajectories using the four FFs. The results of the ARND of water in the parameter region $r < 0.35$ nm are presented in Fig. 6. From Fig. 6, one can find that the distribution of the interface water around the eight kinds of oxygen atoms in the four systems displays a clear distinction. For example, the ARND of water around O32 under the Berger FF is high than that under the Kukol FF. Thus, the oxygen atoms O32 in the former system have a better chance to form H-bond with water than those in the latter system. However, the same reason cannot be extended to explain why the number of H-bond for all other types of oxygen molecule is different among the four systems. The relationship between the ARND of water near O21 and the average number of H-bonds for O21 is a case. As can be seen from Table 2, the average numbers of H-bonds for O21 given from Slipid FF is smaller than that determined by Kukol FF, although the value of the ARND of water near O21 in the former system is greater than that in the latter system. Thus, the average number of H-bonds is not only determined by the ARND of water around the oxygen molecule in spite of its important role in the formation of H-bond. This is because that the ability of a single water molecule to form H-bond with the oxygen molecule of DPPC in the HB parameter area, which is defined as the area within 0.35 nm from the oxygen molecule, differs in different systems. The ability depends on the probability of one water to form HB with DPPC if it exits in the above parameter region.

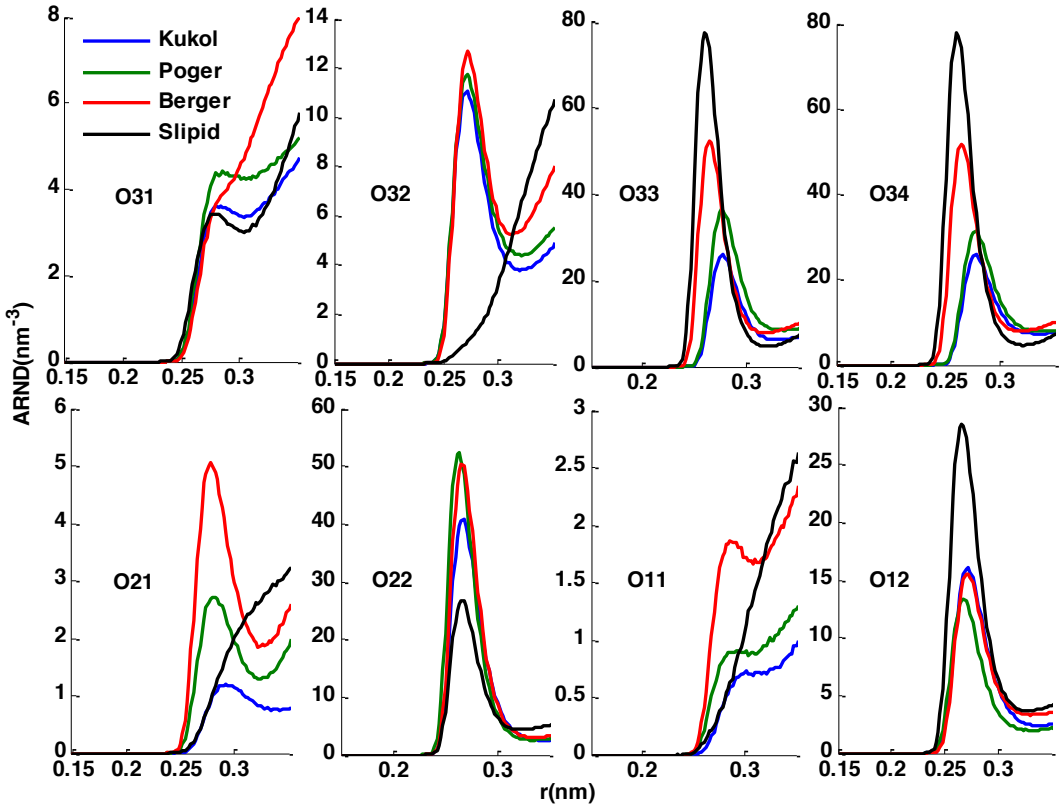

**Fig. 6.** The ARND of water of simulated system using Slipid FF, Berger FF, Poger FF and Kukol FF.

Here, we estimate the probability by using the following formula

$$P_{formHB} = \frac{\sum_t \sum_i N_{tiHB}/N_{iC}}{N_{total}} \qquad (1)$$

where $N_{ic}$ is the times of acquiring data for a time $t$ when a water molecule $i$ stays continuously in the area within 0.35 nm from one certain oxygen molecule of DPPC during the MD simulation, $N_{tiHB}$ is the times that $i$ was observed to form HB with that oxygen molecule during the same time, and $N_{total}$ is the total number of times of all water molecules entering into the above parameter region in a simulation process. This function $P_{formHB}$ can obviously measure the probability that a single water molecule forms HB with a type of oxygen molecule of DPPC. In Fig. 7, the results of $P_{formHB}$ of each simulation system as a function of $N_{total}$ are presented. As shown in Fig. 7, the variety of the value of $N_{total}$ has little effect on the $P_{formHB}$ calculation results. Thus, one can use formula (1) to detect the probability of the formation of a HB between a molecule of water and one type of oxygen molecule of DPPC reliably. It can be seen that the double-bonded oxygen atom in a certain group of one system, compared with the single-bonded one, possesses a stronger ability to form hydrogen bonds with water. The difference of the ability to form HB with water between the double-bonded oxygen atom and the single-bonded oxygen atom, coupled with the fact that the ARND of water around the double-bonded oxygen atom is high than that around the single-bonded oxygen atom, lead to the formation of HB between water and DPPC mainly occurs between the double-bonded oxygen atom and the water molecule, as described above. Now, one can conclude that the distribution of the water around the eight types of the oxygen atom and the ability of a single water molecule to form HB with the oxygen atom of DPPC determine the feature of the hydrogen bond network between water and DPPC.

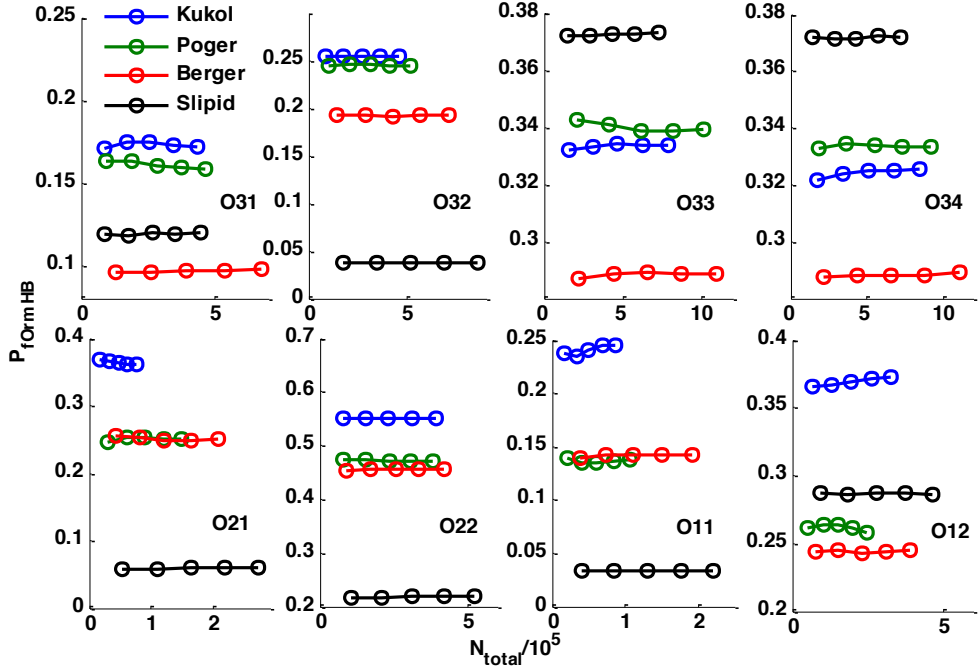

**Fig. 7.** The probability of $P_{formHB}$ of eight types of oxygen atoms from simulations with different FFs.

## IV. CONCLUSIONS

In this paper, we investigated the water-phospholipid interactions at the lipid bilayer/water interfaces under three united-atom FFs (Beger, Kukol and Poger) and an all-atom FF (Slipid) in MD simulations. We found that all these FFs described lipid bilayer well, and the structural properties are in a good agreement with results from experimental measurements, including the mass distributions of different groups positions, and the deuterium order parameters of hydrocarbon chains in the lipid bilayers (DPPC). However, there are evident differences between the water-phospholipid interactions at lipid bilayer/water interfaces under the four FFs. At lipid bilayer/water interfaces, the formation of hydrogen bonds between water molecules and phospholipids are mostly involved with the eight oxygen atoms of hydrophilic heads of DPPC, and more than 80% of these hydrogen bonds are linked to the four double-bonded oxygen atoms in the eight oxygen atoms, namely O33, O34, O22, O12. We compare the results of these hydrogen bonds for the four FFs, and find the proportion of the hydrogen bonds linked to the four double bonded oxygens in all hydrogen bonds of phospholipids with water for the all-atom FF (Slipid) is up to 96% and is much larger than those for three used united-atom FFs, and the averaged numbers of hydrogen bonds for the four double bonded oxygens are evidently different. Further, we compare the energy distribution of hydrogen bonds between the eight oxygen atoms of hydrophilic heads and water and also between different head groups of phospholipids and water for the four FFs. Obviously, in comparison with the Slipid FF, the three used united-atom FFs evidently weaken hydrogen bonds of the phosphate groups of phospholipids with water, but in contrast those hydrogen bonds of the glycerol groups are enlarged. The different interfacial water distributions of the above four systems, combined with the fact that the ability of the interfacial water to form the hydrogen bond with oxygen atoms of the DPPC lipids is different under different FF, cause the occurring of the above phenomenon. The implicit treatment of all aliphatic hydrogen atoms of phospholipids can reduce the great computation costs with respect to all-atom lipid FF, but it causes the change of the distribution of the interface water, and eventually may bring about evident influence on various interacting processes at lipid bilayers/water interfaces, e.g., membrane-protein interactions. The realistic hydrogen-bonding structures at the lipid bilayer/water interfaces remain unknown due to the lack of experimental data, but these differences should be noticed in relevant studies of lipid bilayers, especially when water behaviors in the vicinity of the surface of lipid bilayers are considered. Our present results contribute to a delicate understanding of water behaviors at lipid bilayer/water interfaces.

## ACKNOWLEDGMENTS

This work is supported by National Natural Science Foundation of China (NSFC) Grant Nos. 11675138, 11422542 and 11605151, and the Natural Science Foundation of Jiangsu Province, China Grant No. BK2016021645.

**References:**
[1] J. Milhaud, Biochim. Biophys. Acta **1663**, 19 (2004).
[2] M. L. Berkowitz, D. L. Bostick, and S. Pandit, Chem. Rev. **106**, 1527 (2006).
[3] B. D. Ladbrooke, R. M. Williams, and D. Chapman, Biochim. Biophys. Acta **150**,

333 (1968).

[4] R. P. Rand and V. A. Parsegian, Biochim. Biophys. Acta **988**, 351 (1989).

[5] Z. Arsov, *Long-Range Lipid-Water Interaction as Observed by ATR-FTIR Spectroscopy* (Springer International Publishing, 2015).

[6] J. Fitter, R. E. Lechner, and N. A. Dencher, J. Phys. Chem. B **103**, 8036 (1999).

[7] K. Arnold, L. Pratsch and K. Gawrisch, Biochim. Biophys. Acta. **728**, 121 (1983).

[8] P. T. T. Wong and H. H. Mantsch, Chem. Phys. Lipids **46**, 213 (1988).

[9] W. Pohle, D. R. Gauger, M. Bohl, E. Mrazkova, and P. Hobza, Biopolymers **74**, 27 (2004).

[10] M. L. Berkowitz and R. Vácha, Acc. Chem. Res. **45**, 74 (2012).

[11] W. Hübner and A. Blume, Chem. Phys. Lipids **96**, 99 (1998).

[12] M. Pasenkiewicz-Gierula, Y. Takaoka, H. Miyagawa, K. Kitamura, and A. Kusumi, J. Phys. Chem. A **101**, 3677 (1997).

[13] T. Róg, K. Murzyn and M. Pasenkiewicz-Gierula, Chem. Phys. Lett. **352**, 323 (2002).

[14] C. F. Lopez, S. O. Nielsen, M. L. Klein, and P. B. Moore, J. Phys.Chem. B **108**, 6603 (2004).

[15] S. Y. Bhide and M. L. Berkowitz, J. Chem. Phys. **123**, 224702 (2005).

[16] J. A. Mondal, S. Nihonyanagi, S.Yamaguchi, and T.Tahara, J. Am. Chem. Soc. **134**, 7842 (2012).

[17] O. Berger, O. Edholm, and F. Jahnig, Biophys. J. **72**, 2002 (1997).

[18] A. Kukol, J. Chem. Theory Comput. **5**, 615 (2009).

[19] D. Poger, W. F. Van Gunsteren, and A. E. Mark, J. Comput. Chem **31**, 1117 (2010).

[20] J. B. Klauda, R. M. Venable, J. A. Freites, J. W. O'Connor, D. J. Tobias, C. Mondragon-Ramirez, I. Vorobyov, A. D. Jr. MacKerell, and R. W. Pastor, J. Phys. Chem. B **114**, 7830 (2010).

[21] J. P. M. Jämbeck and A. P. Lyubartsev, J. Phys. Chem. B **116**, 3164 (2012).

[22] C. J. Dickson, B. D. Madej, A. A. Skjevik, R. M. Betz, K. Teigen, I. R. Gould, and R. C.Walker, J. Chem. Theory Comput **10**, 865 (2014).

[23] H. I. Petrache, S. W. Dodd, and M. F. Brown, Biophys. J. **79**, 3172 (2000).

[24] J.-P. Douliez, A. Léonard, and E. J. Dufourc, Biophys. J. **68**, 1727 (1995).

[25] E. Egberts, S. J. Marrink, and H. J. C. Berendsen, Eur Biophys J **22**, 423 (1994).

[26] W. L. Jorgensen and J. Tirado-Rives, J. Am. Chem. Soc **110**, 1657 (1988).

[27] S. -W. Chiu, M. Clark, V. Balaji, S. Subramaniam, H. L. Scott, and E. Jakobsson, Biophys. J. **69**, 1230 (1995).

[28] C. Oostenbrink, A.Villa, A. E. Mark, and W. F. Van Gunsteren, J. Comput. Chem **25**, 1656 (2004).

[29] H. J. C. Berendsen, J. P. M. Postma, W. F. van Gunsteren, and J. Hermans, Interaction Models for Water in Relation to Protein Hydration (Springer Netherlands, 1981).

[30] W. L. Jorgensen, J. Chandrasekhar, J. D. Madura, R. W. Impey, and M. L. Klein, J. Chem. Phys. **79**, 926 (1983).

[31] B. Hess, C. Kutzner, D. van der Spoel, and E. Lindahl, J. Chem. Theory Comput. **4**, 435 (2008).

[32] R. W. Hockney, S. P. Goel, and J. W. Eastwood, J. Comp. Phys. **14**, 148 (1974).

[33] B. Hess, H. Bekker, H. J. C. Berendsen, and J. G. E. M. Fraaije, J. Comp. Chem **18**, 1463 (1997).

[34] H. J. C. Berendsen, J. P. M. Postma, W. F. van Gunsteren, A. DiNola, and R. J. Haak, J. Chem. Phys. **81**, 3684 (1984).

[35] D. M.York, T. A. Darden, and L. G. Pedersen, J. Chem. Phys. **99**, 8345 (1993).

[36] A. Luzar and D. Chandler, Phys. Rev. Lett. **76**, 928 (1996).
[37] F. H. Stillinger and A. Rahman, J. Chem. Phys. **60**, 1545 (1974).
[38] J. Zielkiewicz, J. Chem. Phys. **123**, 104501 (2005).
[39] D. Swiatla-Wojcik, Chem. Phys. **342**, 260 (2007).
[40] A. Kuffel and J. Zielkiewicz, J. Phys. Chem. B **112**, 15503 (2008).
[41] P. Guo, Y. S. Tu, J. R. Yang, C. L. Wang, N. Sheng, and H. P. Fang, Phys. Rev. Lett. **115**, 186101 (2015).
[42] J. F. Nagle and S. Tristram-Nagle, Biochim. Biophys. Acta **1469**, 159 (2000).